\documentclass[preprint,showpacs,amsmath,amssymb]{revtex4}
\usepackage{graphicx}
\usepackage{dcolumn}
\usepackage{epsfig}
\usepackage{bm}
\newcommand{\be}{\begin{eqnarray}}
\newcommand{\en}{\end{eqnarray}}
\newcommand{\bes}{\begin{subequations}}
\newcommand{\ens}{\end{subequations}}
\newcommand{\ben}{\begin{eqnarray*}}
\newcommand{\enn}{\end{eqnarray*}}
\newcommand{\beq}{\begin{equation}}
\newcommand{\enq}{\end{equation}}
\newcommand{\pa}{\partial}

\newcommand{\f}{\frac}

\newcommand{\mc}[1]{\mathcal{#1}}

\newcommand{\bi}{\begin{itemize}}
\newcommand{\ei}{\end{itemize}}

\renewcommand{\t}{\Theta}
\renewcommand{\theta}{\Theta}

\renewcommand{\v}{\varepsilon}
\renewcommand{\o}{\omega}

\begin{document}
\title{Center or Limit Cycle: Renormalization Group as a Probe}
\author{Amartya Sarkar}
\email{amarta345@bose.res.in}
\author{J.K. Bhattacharjee}
\email{jkb@bose.res.in}
\affiliation{Department of Theoretical Sciences, S. N. Bose National Centre for Basic Sciences, Salt lake, Kolkata 700098, India}
\author{Sagar Chakraborty}
\email{sagar@nbi.dk}
\affiliation{NBIA, Niels Bohr Institute, Blegdamsvej 17, 2100 Copenhagen $\O$, Denmark}
\author{Dhruba Banerjee}
\email{dhruba.iacs@gmail.com}
\affiliation{Department of Physics, Jadavpur University, Kolkata 700032, India}

\date{\today}
\begin{abstract}
Based on our studies done on two-dimensional autonomous systems, forced non-autonomous systems and time-delayed systems, we propose a unified methodology --- that uses renormalization group theory --- for finding out existence of periodic solutions in a plethora of nonlinear dynamical systems appearing across disciplines. The technique will be shown to have a non-trivial ability of classifying the solutions into limit cycles and periodic orbits surrounding a center. Moreover, the methodology has a definite advantage over linear stability analysis in analyzing centers.
\end{abstract}
\pacs{05.10.Cc, 47.20.Ky, 02.30.Mv}
\maketitle
\section{Introduction}
The study of nonlinear differential equations \citep{nldjorsmi,nldstr} in a two dimensional dynamical system is of considerable interest to researchers across disciplines. Various methods of obtaining approximate analytic solutions have been formulated over the years like Lindstedt-Poincar$\acute{\textrm{e}}$ method, harmonic balance \textit{etc}.
More recently a number of new methods have been proposed, \textit{e.g.} nonperturbative method \citep{bdelprl,dpolpre}, $\delta$-method \citep{benpur}, homotopy perturbation method \citep{hejnm1}, variational iteration methods \citep{hejnm2} \textit{etc}.
The use of renormalization group (RG) in the analysis of nonlinear dynamical problems \citep{benors} has been pioneered by Goldenfeld and co-workers \cite{chenet,chenet1,chenet2}.
The traditional perturbative theory (for example, the multiple scale method) relies on one's ability to recognize the correct scales.
The use of RG on the direct perturbation expansion, eliminated the necessity of recognizing the correct scales --- the scales emerged automatically on implementation of the RG.
\\
What we show in this paper is how this RG technique can be of use in devising a methodology (described later in this paper) capable of distinguishing between different types of periodic solutions --- centers and limit-cycles --- in two-dimensional autonomous dynamical systems of the general form: $\dot{x}=P(x,y),\,\dot{y}=Q(x,y)$.
The presence of limit cycles in a model facilitates explanation of self-sustained oscillations.
Limit cycles appear in wide variety of modern researches in many fields like quantum physics, chemical physics, biophysics, material sciences, ecology \textit{etc.} --- see {\it e.g.,} \citep{qp,cp,bp,mp,ep} respectively for recent examples.
Consequently, finding out variety of methods \citep{bdelprl,giavia,jheprl} (however, see also \citep{comprl}) for determining limit cycles in nonlinear problems is of current research interest.
These orbits are essentially nonlinear in nature and occur isolated in a phase-space unlike the family of periodic orbits around a center.
We shall also illustrate that this very technique can help one ascertain if a fixed point is a focus or a center.
It is worth mentioning that distinguishing between a center and a focus (known as {\textit {center problem}}) is one of the main and oldest problems in two-dimensional dynamical systems.
Our method of distinguishing focus, center and limit-cycle will be easily shown to be extendable to two-dimensional non-autonomous systems and also to the extremely important class of time-delayed dynamical systems.
Moreover, our technique will yield the correct nature of a fixed point of a nonlinear dynamical system when the linearization about the fixed point gives a completely wrong idea regarding the true nature of the fixed point.
\\
Attempts in solving an ordinary differential equation of the form:
$\ddot{x}+\o^2 x=\v F(x,\dot{x})$,
using a naive expansion, $x(t)=x_0+\v x_1+{\v}^2 x_2+\cdots$, results in breakdown of the perturbation theory at times $t$ such that $\v (t-t_0)>1$ (where $t_0$ is the initial time) due to the presence of secular terms. How does one apply the RG principle to the problem? We begin by observing that a periodic solution can be expressed as a Fourier series with amplitude A and phase $\theta$ of the lowest harmonic, determining the amplitude and phase of the higher order ones. The amplitude and phase are quantities that will {\it flow}. To regularize the perturbation series, RG technique first introduces an arbitrary time $\tau$ with a view to splitting $t-t_0$ as $(t-\tau)+(\tau-t_0)$ and absorbing the terms containing $\tau-t_0$ into the respective {\it renormalized} counterparts $A$ and $\t$ of $A_0$ and $\t_0$. $A_0$ and $\t_0$ are the constants of integration determined at $t_0$. This is completely similar to divergence in field theories where a physical quantity ({\it e.g.}, two point correlation function) diverges as the cutoff $\Lambda\rightarrow\infty$. If we are discussing a physical variable, then the answer has to be finite and while this is achieved in field theory by constructing running coupling constants, it is done for the differential equation by introducing an arbitrary time scale $\tau$ and letting the amplitude and phase depend on $\tau$. At the end of the process one arrives at the RG-flow-equations for $A$ and $\t$:
\be
\f{d A}{d\tau}=f(A,\t);\quad\f{d \t}{d\tau}=g(A,\t).
\label{eq2}
\en
 So, we see that the RG naturally leads to flow equations. In this respect it is akin to the Bogoliubov-Krylov method \cite{nldjorsmi}. But as mentioned earlier, the advantage lies in the fact that RG uses naive perturbation theory; and we do not need to anticipate scales (as in multiple scales method) or make an assumption about slowly varying amplitudes and phases (Bogoliubov-Krylov).\\
For autonomous systems, $f$ and $g$ are generally function of $A$ alone. We propose to use flow equations (1.1)  and (1.2) to differentiate between oscillators which are of the center variety and limit cycles. The center type oscillation consists of a continuous family of closed orbits in phase space, each orbit being determined by its own initial condition. This implies that the amplitude $A$ is fixed once the initial condition is set. This must lead to
\begin{equation}
\frac{dA}{d\tau}=0
\end{equation}
This statement is exact and is not tied to any perturbation theory argument. For the limit cycle on the other hand
\begin{equation}
\frac{dA}{d\tau}=f(A)\label{1.4}
\end{equation}
	and $f(A)$ must be such that the flow has a fixed point. The fixed point has to be stable for the limit cycle to be stable. Also, if $A=0$ is a fixed point of equation (\ref{1.4}), then we have a focus.
\\
This extremely simple prescription, though not proved rigorously, appeals to one's intuition when one notes that $(i)$ $A=0$ means the {\textit {assumed}} periodic solution has zero amplitude and hence hints at focus, $(ii)$ $f^{(A)}=0\,\forall\,A\ge 0$ hints at a family of non-isolated periodic orbits surrounding the fixed point and therefore existence of center is implied, and $(iii)$ vanishing of $dA/d\tau$ at $A=A_i\ne 0$ logically indicates that an isolated periodic orbit of amplitude $A_i$ happens to be surrounding the fixed point.\\
The calculation of  $f(A)$ requires the use of perturbation theory. Application of perturbation theory is possible only if one can locate a center --- this is the basic periodic state. Locating a center can sometimes be straightforward e.g. $\dot{x}_1=x_2,\, \dot{x}_2=-\pa V/\pa x_1$,  where $V$ is a general anharmonic potential: $V= x_1^2/2+\lambda_1 x_1^3/3+\lambda_1 x_1^4/4$. Here $(x_1,x_2 )=(0,0)$ is a linear center around which perturbation theory can be done. Similarly for the Van der Pol oscillator $\dot{x}_1=x_2,\, \dot{x}_2=k\dot{x}_1 (x_1^2-1)+\omega^2 x_1$, the origin is a center for $k=0$. In the Lotka-Volterra model --- $\dot{x}_1=x_1-x_1 x_2,\, \dot{x}_2=-x_2+x_1 x_2$ the origin is a saddle and $(1,1)$ is the center. Shifting the origin to the center is the first step of the process of determining the function $f(A)$. In this case, of course, $f(A)=0$ since the periodic state in the predator-prey model is a center like state.\\
A more complicated situation arises in the Belushov-Zhabotinsky reaction \cite{epslen,lenetl} system. In that case, a transfer of origin to the fixed point will have to be followed by a proper setting of parameters in the problem to make the origin a center which is the starting point of all perturbation theory. This raises the problem that the given dynamical system may not have a relevant parameter, e.g. the well known paradigm for the limit cycle
\be
\dot{z}=(1+i)z-\beta|z|^2 z
\en
	where $z=x+iy$ is the complex variable and  $\beta>0$. The only fixed point is the origin and it is an unstable focus for all $\beta$. We can overcome this difficulty by considering the more general system
\be
\dot{z}=(\alpha_1+i\alpha_2 )z-\beta|z|^2 z
\en
The origin is now a stable focus for $\alpha_1<0$, unstable focus for $\alpha_1>0$ and a center for $\alpha_1=0$. It is this center about which one can set up a perturbation theory.
The perturbative evaluation of $f(A)$ and $g(A)$ consequently involves the following initial steps:
\begin{enumerate}
	\item Find the fixed points of the system and identify linear centers.
\item	If there are no linear centers, extend the parameter space and see if a linear center can be located as the parameters are changed.
\item	For every linear center, thus located, we need to check the existence of a limit cycle by perturbatively constructing $f(A)$ and $g(A)$.
\end{enumerate}
\section{center}
In this section, we take up the study of center. The first example of the nonlinear oscillator will be dealt with in detail --- the subsequent ones will be handled briefly.
\subsection{Unforced Duffing oscillator}
The equation of motion of this damped nonlinear oscillator is
\be
\ddot{x}+k\dot{x}+\omega^2 x+\lambda x^3 = 0 \label{32001}
\en
We notice that a linear center exists for $k=\lambda=0$. Hence the perturbation theory will have to be built around this limit. We expand
\be
x=x_0+kx'_1+\lambda x_1+k^2x'_2+\lambda^2 x_2+k\lambda x''_2+\dots
\label{32002}
\en
Putting Eq.(\ref{32002}) in Eq.(\ref{32001}), we obtain:
\be
\ddot{x}_0+\omega^2 x_0 &=& 0\label{32003}\\
\ddot{x}_1+\omega^2 x_1 &=& -x_0^3\label{32004}\\
\ddot{x'}_1+\omega^2 x'_1 &=& -\dot{x}_0\label{32005}
\en
With the initial condition set as $x(t=0)=A$ and $\dot{x}(t=0)=0$, we write the solution of Eq.\eqref{32003} as
\beq
x_0=A\cos \omega t\label{32006}
\enq
We note that $x_0$ picks up the initial condition and hence $x_i(t=0)=\dot{x}_i(t=0)=0$ for all $i\geq 1$. Accordingly, Eq.(\ref{32004}) and Eq.(\ref{32005}) now read:
\be
\ddot{x}_1+\omega^2 x_1 &=& -\f{A^3}{4}\left(\cos 3\omega t + 3 \cos\omega t\right)\label{32007}\\
\ddot{x'}_1+\omega^2 x'_1 &=& \omega A\sin\omega t\label{32008}
\en
giving rise to the following solutions respectively
\be
x_1 &=& -\f{3A^3}{8\omega}t\sin\omega t+\f{A^3}{32\omega^2}(\cos 3\omega t-\cos\omega t)\label{32009}\\
x'_1 &=& -\f{A}{2}t\cos\omega t+\f{A}{2\omega}\sin\omega t\label{32010}
\en
keeping in mind the initial conditions. At this order, the displacement of the oscillator is
\be
x(t) &=& A\cos \omega t-\f{3\lambda A^3}{8\omega}t\sin\omega t+\f{\lambda A^3}{32\omega^2}(\cos 3\omega t-\cos\omega t)-\f{kA}{2}t\cos\omega t+\f{kA}{2\omega}\sin\omega t\nonumber\\
&=&A\cos \omega t-\f{3\lambda A^3}{8\omega}(t-\tau+\tau)\sin\omega t+\f{\lambda A^3}{32\omega^2}(\cos 3\omega t-\cos\omega t)\nonumber\\ &&-\f{kA}{2}(t-\tau+\tau)\cos\omega t+\f{kA}{2\omega}\sin\omega t\label{32011}
\en
where we have split the interval $0$ to $t$ as $0$ to $\tau$ and $\tau$ to $t$. To remove the divergences, we introduce the renormalization constants $\mc{Z}_1(0,\tau)$ and $\mc{Z}_2(0,\tau)$ as
\bes
\beq
A = A(t=0)=A(\tau)\mc{Z}_1(0,\tau)\label{32012a}
\enq
\beq
0 = \theta(t=0)=\theta(\tau)+\mc{Z}_2(0,\tau)\label{32012b}
\enq
\ens
The renormalization constants have the expansion
\bes
\beq
\mc{Z}_1(0,\tau)=1+a_1\lambda +a'_1k+\dots\label{32013a}
\enq
\beq
\mc{Z}_2(0,\tau)=b_1\lambda+b'_1 k+\dots\phantom{uuu} \label{32013b}
\enq
\ens
so that the constants $a_i$ and $b_i$ can be  chosen order by order to remove divergences at each order.
In terms of $A(\tau)$ and $\theta(\tau)$, we can write Eq.\eqref{32011} as
\be
x(t) &=& A(\tau)\left[1+a_1\lambda+a'_1k\right]\cos(\omega t+\theta(\tau)+b_1\lambda+b'_1k)\nonumber\\
&&-\f{3\lambda A^3}{8\omega}(t-\tau+\tau)\sin(\omega t+\theta)+\f{\lambda A^3}{32\omega^2}(\cos 3(\omega t+\theta)-\cos(\omega t+\theta))\nonumber\\
&&-\f{kA}{2}(t-\tau+\tau)\cos(\omega t+\theta)+\f{kA}{2\omega}\sin(\omega t+\theta)\nonumber\\
&=& A(\tau)\cos(\omega t+\theta)+(a_1\lambda +a'_1\lambda)A(\tau)\cos(\omega t+\theta)-(b_1\lambda+b'_1 k)A(\tau)\sin(\omega t+\theta)\nonumber\\
&&-\f{3\lambda A^3}{8\omega}(t-\tau+\tau)\sin(\omega t+\theta)+\f{\lambda A^3}{32\omega^2}(\cos 3(\omega t+\theta)\nonumber\\
&&-\cos(\omega t+\theta))-\f{kA}{2}(t-\tau+\tau)\cos(\omega t+\theta)+\f{kA}{2\omega}\sin(\omega t+\theta)\label{32014}
\en
correct to $\mc{O}(\lambda)$ and $\mc{O}(k)$. We chose $a'_1=\f{kA\tau}{2}$, $a_1=0$, $b'_1=0$ and $b_1=-\f{3\lambda}{8\omega}\tau$ to write Eq.\eqref{32015} as
\be
x(t,\tau) &=& A(\tau)\cos(\omega t+\theta)-\f{3\lambda A^3}{8\omega}(t-\tau)\sin(\omega t+\theta)+\f{\lambda A^3}{32\omega^2}(\cos 3(\omega t+\theta)-\cos(\omega t+\theta))\nonumber\\
&&-\f{kA}{2}(t-\tau)\cos(\omega t+\theta)+\f{kA}{2\omega}\sin(\omega t+\theta)\label{32015}
\en
\\
We now impose the condition that $x(t)$ has to be independent of $\tau$ i.e. $\dfrac{dx}{d\tau}=0$ and this yields (to the lowest order)
\bes
\beq
\f{dA}{d\tau}=-\f{kA}{2}\label{32016a}
\enq
\beq
\f{d\theta}{d\tau}=\f{3\lambda A^2}{8\omega}\label{32016b}
\enq
\ens
integrating to $A=A_0 e^{-k\tau/2}$ and $\theta=\theta_0+\f{3\lambda A^2}{8\omega}\tau$. Final removal of $\tau$ requires setting $\tau=t$ and then we have
\be
x(t) &=& A_0 e^{-kt/2}\cos\left[\left(\omega+\f{3\lambda A^2}{8\omega}\right)t+\theta_0\right]\nonumber\\&&+\f{\lambda A_0^3}{32\omega^2}(\cos 3(\omega t+\theta_0)-\cos(\omega t+\theta_0))+\f{kA_0}{2\omega}\sin(\omega t+\theta_0)\label{32017}
\en
\\
For $k=0$, we have the conservative anharmonic oscillator
\beq
\ddot{x}+\omega^2 x+\lambda x^3 = 0\label{32018}
\enq
for which the fixed point $(0,0)$ in the $x-\dot{x}$ plane ({\it i.e.} $x-y$ plane) is a center and then as expected
\beq
\f{dA}{d\tau}=0\label{32019}
\enq
with
\be
x(t)=A_0\cos\Omega t+\f{\lambda A_0^3}{32\Omega^2}\left[\cos 3\Omega t-\cos\Omega t\right]+\mc{O}(\lambda^2)\label{32020}
\en
where
\beq
\Omega=\omega+\f{3\lambda A_0^2}{8\omega}+\mc{O}(\lambda^2)\label{32021}
\enq
The standard results \cite{jkbms} for the oscillator have, thus, been
correctly captured and we find that the emergence of $x = \dot{x} = 0$ as a center is confirmed by the fact that $dA/d\tau = 0$.
It is worth pointing out a couple of features in the calculation. The diverging terms in the perturbative solution come from the lowest harmonic $sine$ and $cosine$ terms on the right hand side (inhomogeneous term). The $sine$ term is responsible for the amplitude flow ${dA}/{d\tau}$ and the $cosine$ term is responsible for the phase flow. To derive the amplitude equation, this is what we need to concentrate on. Keeping this in mind, if we examine the structure of higher order terms, we find that a $sine$ term is never generated on the right hand side and hence ${dA}/{d\tau}=0$ at all orders. This question of lowest harmonic $sine$ and $cosine$ terms is quite general and we can subsequently use this to write down the flow equation by inspection. It should be borne in mind that although the lowest harmonic is unnecessary for writing down the flow equation, it is imperative to have all the relevant harmonics for writing down the actual solution $x(t)$ at any order. For the anharmonic oscillator of Eq.\eqref{32018}, the phase space trajectory is
\beq
\dot{x}^2+\omega^2 x^2+\frac{\lambda}{2}x^4=\textrm{constant}=A_0^2+\f{\lambda}{2}A_0^4\label{32022}
\enq
where $(A_0,0)$ is the initial condition for the trajectory. It is straightforward to check that Eq.\eqref{32020} and Eq.\eqref{32021} are in exact agreement with Eq.\eqref{32022} to $\mc{O}(\lambda)$. Thus, perturbatively the correct phase portrait is obtained as it should. But as is well known, the perturbation series is not convergent and things are bound to get worse as we go to higher values of $\lambda$. This is a problem with all perturbative approaches.
\subsection{Lotka-Volterra System}
Having explained in detail the case of the anharmonic oscillator, we now turn to the predator-prey model which is also known to have oscillatory trajectories. The prey population is $x$ and the predator population is $y$, with the dynamics given by
\begin{subequations}\label{32023}
\begin{equation}
\frac{dx}{dt} = x-xy\phantom{u}\label{32023a}
\end{equation}
\begin{equation}
\frac{dy}{dt} = -y+xy \label{32023b}
\end{equation}
\end{subequations}
The origin turns out to be a saddle and there is a center at $(1,1)$. We shift the origin to the center --- a procedure which will be followed regularly in our perturbative calculation of the periodic trajectories. Accordingly we define
\be
x &=& X+1\nonumber\\
y &=& Y+1\nonumber
\en
and write the system as:
\bes \label{32024}
\beq
\dot{X}=-Y-XY \label{32024a}
\enq
\beq
\dot{Y}=X+XY\phantom{u} \label{32024b}
\enq
\ens
The perturbative theory proceeds by imagining the existence of a coupling constant $\lambda$ in terms of which we have
\bes \label{32025}
\beq
\dot{X}=-Y-\lambda XY \label{32025a}
\enq
\beq
\dot{Y}=X+\lambda XY\phantom{u} \label{32025b}
\enq
\ens
We expand $X$ and $Y$ as $X= X_0+\lambda X_1+\lambda^2 X_2+\cdots$ and $Y= Y_0+\lambda Y_1+\lambda^2 Y_2+\cdots$ respectively to subsequently arrive at
\bes
\be
\dot{X}_0 &=& -Y_0\label{32026a}\\
\dot{Y}_0 &=& X_0\label{32026b}
\en
\be
\dot{X}_1 &=& -Y_1-X_0 Y_0\label{32026c}\\
\dot{Y}_1 &=& X_1+X_0 Y_0\label{32026d}
\en
\be
\dot{X}_2 &=& -Y_2-(X_0 Y_1+Y_0 X_1)\label{32026e}\\
\dot{Y}_2 &=& X_2+(X_0 Y_1+Y_0 X_1)\label{32026f}
\en
\ens
and so on. Clearly $X_0=A\cos t$ (initial condition being $x=A$, $\dot{x}=0$ at $t=0$) and $Y_0=A\sin t$. At the next order
\beq
\ddot{X}_1+X_1=-(L_0+\dot{L}_0)\label{32027}
\enq
where $L_0=X_0Y_0$ and hence
\beq
\ddot{X}_1+X_1=-\left(\f{A^2}{2}\sin 2t+A^2\cos 2t\right)\label{32028}
\enq
There are no resonating terms on the right hand side and thus
\beq
X_1=\f{A^2}{6}\left(\sin 2t-\sin t\right)+\f{A^2}{3}\left(\cos 2t-\cos t\right)\label{32029}
\enq
keeping the initial condition $X_1(t=0)=\dot{X}_1(t=0)=0$ in mind.
 
\begin{figure}
\begin{center}
\includegraphics[width=10 cm]{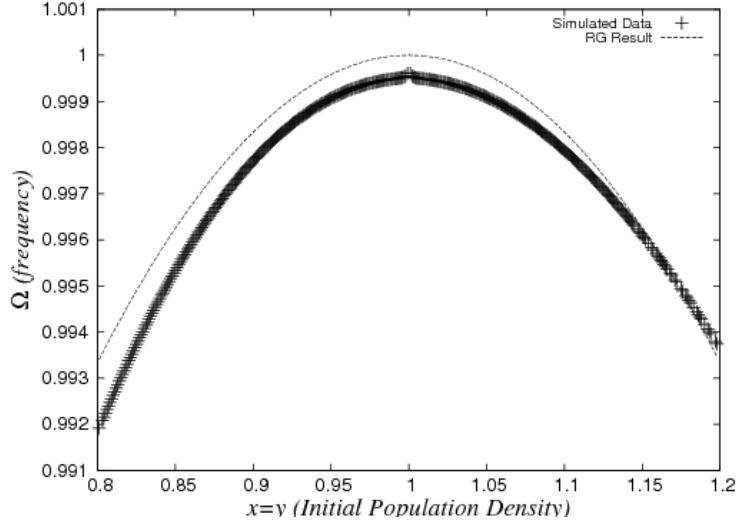}
\end{center}
\caption{The initial population density is plotted against the corresponding frequency. The figure makes it clear the RG calculation is reasonably correct.}
\label{f:1}
\end{figure}

At the succeeding order
\beq
\ddot{X}_2+X_2=-(L_{10}+\dot{L}_{10})\label{32030}
\enq
where $L_{10}=X_0 Y_1+Y_0 X_1$. We see immediately that the resonant term on the right hand side of Eq.\eqref{32030} is ${A^3}\cos t/12$ and in keeping with our previous discussion, it follows that
\bes
\beq
\f{dA}{d\tau}=0\label{32031a}
\enq
\beq
\f{d\theta}{d\tau}=-\f{A^2}{12}\label{32031b}
\enq
\ens
This gives an amplitude dependent frequency of
\beq
\Omega=1-\f{A^2}{12}\label{32032}
\enq
which has been numerically verified as can be seen from Fig. {\ref{f:1}}.
\subsection{A Lienard system}
A more interesting example is the dynamical system
\bes \label{32033}
\beq
\dot{x} = y\phantom{-x^2+y\left(1+x+\mu\right)uu}\label{32033a}
\enq
\beq
\dot{y} = -x-x^2+y\left(1+x+\mu\right)\label{32033b}
\enq
\ens
The two fixed points of the above system are at $(0,0)$ and $(-1,0)$. The former is a center and the latter a saddle. The oscillatory orbit needs to be investigated around $(0,0)$. The above system is a second order differential equation of the Lienard variety $\ddot{x}+\dot{x}F(x)+G(x)=0$, where $F(x)$ is a linear function $\alpha+\beta x$ and $G(x)=x+\lambda x^2$. We redefine $\alpha$ and $\beta$ to write
\beq
\ddot{x}-k\dot{x}\left(1+x+\mu\right)+x+\lambda x^2 = 0\label{32034}
\enq
Clearly Eq.\eqref{32033} is obtained for $k=1$ and $\lambda=1$. The perturbation theory has to proceed around the linear center which is at $k=\lambda=0$. Accordingly, we expand
\beq
x=x_0+kx_1+\lambda x'_1+k^2x_2+\lambda x'_2+k\lambda x''_2+\dots \label{32035}
\enq
At different orders we have
\bes
\beq
\mc{O}(k^0\lambda^0):\quad\quad\ddot{x}_0+x_0=0\phantom{uuuuuuuuuuuuuuuuuuuuuu} \label{32036a}
\enq
\beq
\mc{O}(k^1\lambda^0):\quad\quad\ddot{x}_1+x_1=\dot{x}_0(1+\mu)+x_0 \dot{x}_0\phantom{uuuuuuuuuuu}\label{32036b}
\enq
\beq
\mc{O}(k^0\lambda^1 ):\quad\quad\ddot{x}'_1+x'_1=x_0^2\phantom{uuuuuuuuuuuuuuuuuuuuluu} \label{32036c}
\enq
\beq
\mc{O}(k^2\lambda^0):\quad\quad\ddot{x}_2+x_2= \dot{x}_1(1+\mu)+x_1\dot{x}_0+x_0\dot{x}_1\phantom{uuuuuu}\label{32036d}
\enq
\beq
\mc{O}(k^0\lambda^2):\quad\quad\ddot{x}'_2+x'_2=-2x_0x'_1\phantom{uuuuuuuuuuuuuuuuuuu}\label{32036e}
\enq
\beq
\mc{O}(k^1\lambda^1):\quad\quad\ddot{x}''_2+x''_2= x'_1(1+\mu)+x_0\dot{x'}_1+x'_1\dot{x}_0-2x_0x_1\label{32036f}
\enq
\ens
With the initial condition $x=A$, $\dot{x}=0$ at $t=0$,
\bes
\beq
x_0=A\cos t \phantom{uuuuuuuuuuuuuuuuuuuuuuuuuuuuuu}\label{32037a}
\enq
\beq
x_1=(1+\mu)\f{A}{2}\left(t\cos t-\sin t\right)+\f{A^2}{6}\left(\sin 2t-2\sin t\right) \label{32037b}
\enq
\beq
x'_1=-\f{A^2}{2}+\f{A^2}{6}\cos 2t + \f{A^2}{3}\cos t \phantom{uuuuuuuuuuuuuuu}\label{32037c}
\enq
\ens
The corresponding flow equations at this order are
\bes
\beq
\f{dA}{d\tau}=\left(1+\mu\right)\label{32038a}
\enq
\beq
\f{d\theta}{d\tau} = 0\phantom{uuuuu}\label{32038b}
\enq
\ens
If we were to have the possibility of a center, then clearly $\mu=-1$. In this section we focus on the potential center and work with $\mu=-1$. At the next order, the flow becomes
\bes
\beq
\f{dA}{d\tau}=-k\lambda\f{A^3}{8}\phantom{uuuuuuuuu}\label{32039a}
\enq
\beq
\f{d\theta}{d\tau}=-\f{5}{12}\lambda^2 A^2-\f{1}{24}k^2A^2\label{32039b}
\enq
\ens
The origin is a focus if both $k$ and $\lambda$ are not zero, but a center if either or both of $k$ and $\lambda$ vanish.\\
The above example is an interesting example of the usefulness of increasing the space of parameters in a dynamical system. Starting with Eq.\eqref{32033}, one would not have access to the different possibilities that we have been finding - e.g. the competition between the center and the focus. This is made possible by introducing the two parameters $k$ and $\lambda$. The general system has been treated in \cite{dhrjkb}.
\section{Limit Cycle}
\subsection{Van der Pol oscillator}
In this section, we consider the question of the limit cycle and we introduce the RG flow by recalling the calculations of Chen et al\citep{chenet2} for the Van der Pol oscillator. This system is represented by
\beq
\ddot{x}+\epsilon\dot{x}\left(x^2-1\right)+\omega^2 x=0 \label{33001}
\enq
When looked at as the second order dynamical system $\dot{x}=y$, $\dot{y}=-\epsilon y(x^2-1)-\omega^2x$, there is a fixed point at the origin which is a stable focus for $\epsilon<0$ and unstable focus for $\epsilon>0$. The fixed point is a center for $\epsilon=0$ and we base the perturbation expansion around $\epsilon=0$, expanding $x$ as
\beq
x(t)=x_0(t)+\epsilon x_1(t)+\epsilon^2 x_2(t)+\dots \label{33002}
\enq
At different orders of $\epsilon$,
\be
\ddot{x}_0+\omega^2 x_0 &=& 0 \label{33003}\\
\ddot{x}_1+\omega^2 x_1 &=& -\dot{x}_0(x_0^2-1)\label{33004}
\en
We work with initial condition $x =A_0$ at $t=0$ and $\dot{x}=0$ at $t=0$. Keeping this in mind, we arrive at:
\be
x_0 &=& A_0\cos \omega t \label{33005}\\
x_1 &=& \f{1}{2}\left(A_0-\f{A_0^3}{4}\right)t\cos\omega t-\f{A_0^3}{32\omega^2}\left(\sin 3\omega t-3\sin \omega t\right)\label{33006}
\en
leading to
\be
x=A_0\cos\omega t+\epsilon\left[\f{1}{2}\left(A_0-\f{A_0^3}{4}\right)t\cos\omega t-\f{A_0^3}{32\omega^2}\left(\sin 3\omega t-3\sin \omega t\right)\right]\label{33007}
\en
As before we split the interval $0$ to $t$ as $0$ to $\tau$ and $\tau$ to $t$, define the renormalization constants $\mc{Z}_1$ and $\mc{Z}_2$ by the relation
\bes
\beq
A =A(\tau)\mc{Z}_1(0,\tau)\phantom{uuuuuuuuuuu}\label{33008a}
\enq
\beq
0 = \theta(t=0)=\theta(\tau)+\mc{Z}_2(0,\tau)\label{33008b}
\enq
\ens
The renormalization constants $\mc{Z}_1$ and $\mc{Z}_2$ can be expanded as
\bes
\beq
\mc{Z}_1(0,\tau)=1+\alpha_1\epsilon+\alpha_2\epsilon^2+\dots\label{33009a}
\enq
\beq
\mc{Z}_2(0,\tau)=\beta_1\epsilon+\beta_2\epsilon^2+\dots\phantom{uuu} \label{33009b}
\enq
\ens
To $\mc{O}(\epsilon)$, we now have
\be
x(t) &=& A(1+\alpha_1\epsilon+\alpha_2\epsilon^2+\dots)\left[\cos(\omega t+\theta)-(\beta_1\epsilon+\beta_2\epsilon^2+\dots)\sin(\omega t+\theta)\right]\nonumber\\&&+\f{\epsilon}{2}\left(A_0-\f{A_0^3}{4}\right)(t-\tau+\tau)\cos(\omega t+\theta)-\f{\epsilon A_0^3}{32\omega^2}\left(\sin 3\omega t-3\sin \omega t\right)\label{33010}
\en
We choose
\be
\alpha_1 &=& -\f{\epsilon}{2}\left(A_0-\f{A_0^3}{4}\right)\tau \label{33011}\\
\beta_1 &=& 0 \label{33012}
\en
to remove divergence from the \emph{past}. We are now left with
\be
x(t) = A \cos(\omega t+\theta)+\f{\epsilon}{2}\left(A_0-\f{A_0^3}{4}\right)(t-\tau)\cos(\omega t+\theta)-\f{\epsilon A_0^3}{32\omega^2}\left(\sin 3\omega t-3\sin \omega t\right)\label{33013}
\en
We now impose the condition that ${dx}/{d\tau}=0$ since $\tau$ is an arbitrary time and $x(t)$ cannot depend on where one puts the initial condition. This leads to
\be
\f{dA}{d\tau} &=& \f{\epsilon}{2}\left(A_0-\f{A_0^3}{4}\right)\label{33014}\\
\f{d\theta}{d\tau} &=& 0\label{33015}
\en
The remaining $\tau$-dependence in $x(t)$ is removed by setting $\tau = t$ and thus
\be
x(t) =  A \cos(\omega t+\theta)-\f{\epsilon A_0^3}{32\omega^2}\left(\sin 3\omega t-3\sin \omega t\right)\label{33016}
\en
The flow equation has a stable fixed point at $A^2=4$ and this gives the usual Van der Pol limit cycle of radius 2 for small $\epsilon$.
\subsection{Lienard equation}
We now return to the Lienard equation of Eq.\eqref{32034} and consider what happens at the second order if $\mu+1\neq 0$. We find in a manner identical to that outlined above,
\be
\f{dA}{d\tau} &=& (\mu+1)\f{kA}{2}-k\lambda\f{A^3}{8}\label{33017}\\
\f{d\theta}{d\tau} &=& (\mu+1)-\f{5}{12}\lambda^2 A^2-\f{1}{24}k^2A^2 \label{33018}
\en
There is a limit cycle in the system if $k\neq 0$ and $\lambda\neq 0$; and, $\mu+1$ and $\lambda$ have the same sign. The cycle is stable if $k$ also has the same sign as $\mu + 1$ and unstable otherwise. Opposite signs of $k$ and $\lambda$ do not allow for the existence of a stable limit cycle ---  a fact easily ascertained by numerical experiments. The limit cycle for $k\neq 0$, $\lambda\neq 0$ is obtained by simultaneous relaxing of the conditions on $F(x)$ and $G(x)$ in the Lienard system --- the conditions being $F(x)$ being either odd (\emph{center}) or even (\emph{limit cycle}) with $G(x)$ odd. In our case both $F(x)$ and $G(x)$ are of mixed parity.
\subsection{Glycolytic oscillator}
We now turn to another example which clearly illustrates the use of shifting of origin and determination of the locus of Hopf bifurcation points to set up the perturbation theory and locate the limit cycle. This example is drawn from biology and the subject is glycolysis\citep{golnat,prldsr}. The simplest mathematical model is that of Selkov\citep{sel} and is a 2-dimensional system. The variable $x$ is the concentration of ADP (adenosine diphosphate) and $y$ that of F6P (fructose-6-phosphate). The dynamics is given by
\be
\dot{x} &=& -x+(a+x^2)y\label{33019}\\
\dot{y} &=& b-(a+x^2)y\label{33020}
\en
where '$b$' is the rate of fructose production by the substrate and '$a$' is the rate at which fructose decomposes (converts to ADP). It should be noted that the presence of ADP catalyzes this conversion and hence '$a$' is augmented to $a+x^2$. The fixed point of the system is at
\be
x=b,\quad\quad y={b}/{(a+b^2)} \label{33021}
\en

\begin{figure}
\begin{center}
\includegraphics[width=10 cm]{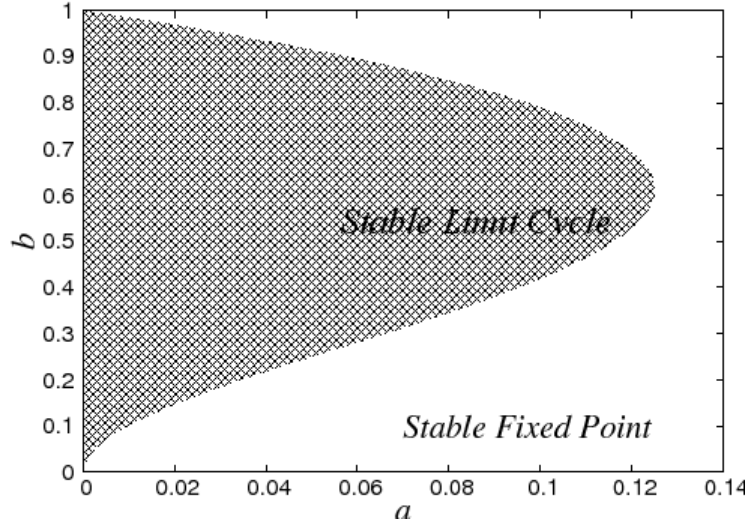}
\end{center}
\caption{The curve: $2a=\sqrt{1+8b^2}-(1+2b^2)$ separates the figure into shaded and unshaded regions. If parameters are in the shaded region, one gets limit cycle (and unstable focus) while unshaded region corresponds to parameters giving rise to stable focus. Linear stability analysis predict centers for parameters on the curve. These centers are, however, not {\it non-linear} centers.}
\label{f:2}
\end{figure}

\begin{figure}
\begin{center}
\includegraphics[width=10 cm]{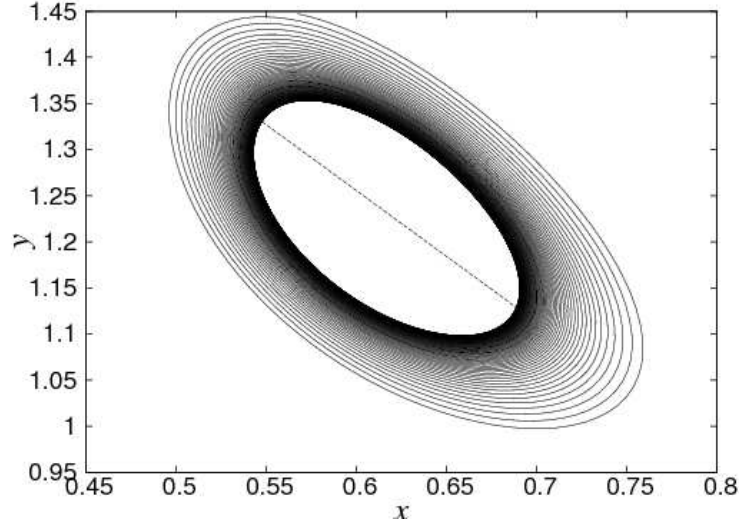}
\end{center}
\caption{Limit cycle in glycolytic oscillator for $a = 0.124$, $b = \sqrt{0.375}$ and $\delta = 0.001$.}
\label{f:3}
\end{figure}

The fixed point is a stable focus for a certain parameter range and an unstable focus for certain others. The crossover from stable to unstable focus occurs on the boundary curve which is a locus of points in the $a$-$b$ plane where a Hopf bifurcation occurs i.e. the fixed point for those values of $(a,b)$ is a center. The curve is given by $2a=\sqrt{1+8b^2}-(1+2b^2)$ and is shown in Fig. {\ref{f:2}}.
For points in the shaded region the fixed point is an unstable focus and for these values of $(a,b)$ a limit cycle can be shown to exist by invoking Poincar$\acute{\textrm{e}}$-Bendixson theorem. We shift the fixed point to the origin and use the new coordinates $X$, $Y$ given by
\be
x &=& b+X \label{33022}\\
y &=& \f{b}{a+b^2}+Y \label{33023}
\en
To use perturbation theory, we chose $(a,b)$ close to the boundary. Setting $b=\sqrt{{3}/{8}}$ (the turning point of the curve), we take $a={1}/{8}-\delta$ to consider a point inside the boundary but close to it. Clearly, $\delta$ is small and positive. To $\mc{O}(\delta)$, the equation of motion reads
\be
\dot{X} &=& \f{1}{2}(X+Y)+\mc{L}(X,Y) \label{33024}\\
\dot{Y} &=& -\f{3}{2}X-\f{Y}{2} - \mc{L}(X,Y) \label{33025}
\en
where
\beq
\mc{L}(X,Y)=\delta(3X-Y)+\sqrt{\dfrac{3}{8}}X(X+Y)+X^2Y \label{33026}
\enq
We note that Eqs. \eqref{33024} and \eqref{33025} combine to give the oscillator,
\beq
\ddot{X}+\f{X}{2}=\dot{\mc{L}} \label{33027}
\enq
$\mc{L}$ has to be expanded in amplitude and the parameter $\delta$. The amplitude will emerge as $\delta^{1/2}$ for small $\delta$. At the zeroth order
\be
X_0 &=& A\cos\left(\f{t}{\sqrt{2}}+\theta\right) \label{33028}\\
Y_0 &=& \sqrt{3}\cos\left(\f{t}{\sqrt{2}}+\theta+\pi-\tan^{-1} \sqrt{2}\right) \label{33029}
\en
The frequency is ${1}/{\sqrt{2}}$ and the axis is tilted at an angle $\pi-\tan^{-1} \sqrt{2}$ to the $X$-axis. The amplitude A is found to from the flow which at the lowest order gives
\beq
\f{dA}{d\tau}=2\delta A -\f{3A^3}{8} \label{33030}
\enq
The frequency changes from the zeroth order value of $\dfrac{1}{\sqrt{2}}$ according to the flow
\beq
\f{d\theta}{d\tau}=-\f{\delta}{\sqrt{2}}+\f{A^2}{4\sqrt{2}} \label{33031}
\enq
The stable fixed point $A^2 =16\delta/3$ gives us the size of the limit cycle for $\delta\ll 1$. A typical small-$\delta$ orbit is shown in Fig. {\ref{f:3}} and bears out the correctness of the above flow. This technique can also be used to probe limit cycles in the more complicated model of Cera et al. \cite{pnascer}.
\subsection{Belushov-Zhabotinsky reaction}
An identical approach is effective for Belushov-Zhabotinsky reaction. A recent version \citep{epslen,lenetl} of that reaction uses a two variable system (chlorine dioxide-iodine-malonic acid reaction)
\be
\dot{x} &=& a - x -\f{4xy}{1+x^2} \label{33032}\\
\dot{y} &=& bx\left(1-\f{y}{1+x^2}\right) \label{33033}
\en
where the variable $x$ and $y$ are the concentrations of the intermediaries $I^-$ and $ClO_2^-$ which vary on a much faster time scale than $ClO_2$, $I_2$ and $Malonic$ $acid$. The constants '$a$' and '$b$' are parameters which depend on the rate constants and the approximately constant concentrations of the other reactants. We note that there is one fixed point $x={a}/{5}$ and $y=1+x^2=1+{a^2}/{25}$. Our first step is to shift the origin to $({a}/{5},1+{a^2}/{25})$ i.e. use the variables
\be
x &=& X+\f{a}{5} \label{33034}\\
y &=& Y +1+\frac{a^2}{25} \label{33035}
\en
The linear stability analysis of the resulting system about the fixed point $X=Y=0$ shows that it is a center for $b=b_c$ given by
\beq
b_c = \f{3a}{5}-\f{25}{a} \label{33036}
\enq
The origin is an unstable focus for $b<b_c$ and stable for $b>b_c$. We pick a value of `$a$' and choose $b=b_c-\delta$, where $\delta\ll b_c$. One carries out a perturbation analysis for the variables $X$ and $Y$ by assuming that the amplitude is small for small $\delta$. The amplitude flow works out to be
\beq
\f{dA}{d\tau} = -\f{a}{5}\delta\Omega A + \f{\Omega A^3}{\left(1+\f{a^2}{25}\right)^2}\left[\f{3a^4}{125}-3a^2-315+\f{1875}{a^2}\right] \label{33037}
\enq
where $\Omega^2 = a\left(1+\f{a^2}{25}\right)\left(\f{3a}{5}-\f{25}{a}\right)$. The limit cycle exists for positive values of $\delta$. It is apparent that as we measure the value of '$a$' for which limit cycles can exist, there is a cyclic-fold bifurcation at $a = a_c\simeq\sqrt{191.43}$ --- obtained by setting the expression inside square bracket to zero.
\subsection{Koch-Meinhardt reaction diffusion system}
\begin{figure}
\begin{center}
\includegraphics[width=10cm]{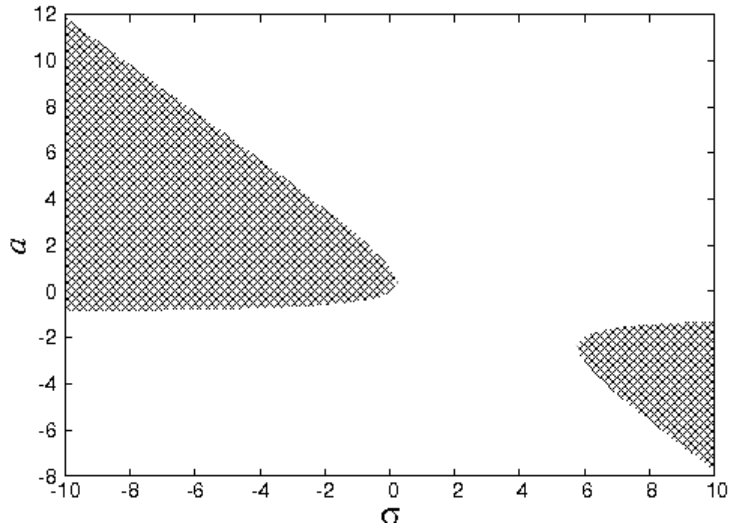}
\end{center}
\caption{$a-\sigma$ parameter space. The shaded regions are where limit cycle solutions can occur.}
\label{f:4}
\end{figure}

Similar considerations apply to a model which is popular for the generation of Turing patterns. This is the Koch-Meinhardt reaction diffusion system \citep{kocmei} and for our present purposes only the reaction part of it is relevant. The variables $x$ and $y$ are the number densities of two species which are responsible for the pigments in the pattern and satisfy the reaction dynamics
\be
\dot{x} &=& -x+\f{x^2}{y}+\sigma \label{33038}\\
\dot{y} &=& -y+x^2\label{33039}
\en
The slowly diffusing pigment $(x)$ is auto-catalytic and also promotes the growth of the antagonistic fast diffusing component $(y)$. The rate of growth of $x$ from the environment is $\sigma$. The fixed point of the above system is $x = 1+\sigma$, $y=(1+\sigma)^2$. As we have made clear, the first step involves shifting the origin to $(1+\sigma,(1+\sigma)^2)$, thus we define
\be
x &=& X + (1+\sigma) \label{33040}\\
y &=& Y +(1+\sigma)^2\label{33041}
\en
In terms of the new variables $X$ and $Y$, the dynamics is

\be
\dot{X} &=& -X + \f{2(1+\sigma)X-Y+X^2}{(1+\sigma)^2+Y} \label{33042}\\
\dot{Y} &=& 2(1+\sigma)X-Y+X^2 \label{33043}
\en
In perturbation theory, we are interested in the small amplitude oscillators around the origin and hence we can expand the denominator in Eq.\eqref{33042} to write
\be
\dot{X} &=& -X+\f{2}{1+\sigma}\left[X-\f{Y}{2(1+\sigma)}+\f{X^2}{2(1+\sigma)}\right]\left[1-\f{Y}{(1+\sigma)^2}+\f{Y^2}{(1+\sigma)^4}+\dots\right]\nonumber\\
&=& X\f{1-\sigma}{1+\sigma}-\f{Y}{(1+\sigma)^2}+\f{X^2}{2(1+\sigma)^2}-\f{2XY}{(1+\sigma)^3}+\f{Y^2}{(1+\sigma)^4}+\f{2Y^2X}{(1+\sigma)^5}\nonumber\\ &&-\f{Y^3}{(1+\sigma)^6}+\dots \label{33044}
\en
The eigenvalues $\lambda$ of a linear stability analysis around the center are found from
\beq
\left(\lambda-\f{1-\sigma}{1+\sigma}\right)\left(\lambda+1\right)+\f{2}{1+\sigma} = 0 \label{33045}
\enq
and are seen to be
\beq
\lambda = -\f{\sigma}{1+\sigma}\pm\sqrt{\f{\sigma^2}{(1+\sigma)^2}-1} \label{33046}
\enq
For no positive $\sigma$, can the origin be an unstable focus and hence it would seem there can be no limit cycle in the system.
\\
However at this point, we generalize this problem by introducing a decay constant `$a$', so that the system (Eqs \eqref{33040} and \eqref{33041}) becomes
\be
\dot{x} &=& -x+\f{x^2}{y}+\sigma \label{33047}\\
\dot{y} &=& -ay+x^2\label{33048}
\en
The fixed point is now at $a+\sigma$, $y=\dfrac{(a+\sigma)^2}{a}$  and using a coordinate system $(\mc{X},\mc{Y})$ centered at the fixed point, we have
\be
\dot{\mc{X}} &=& -\mc{X} + \f{2(a+\sigma)\mc{X}-a\mc{Y}+\mc{X}^2}{\dfrac{(a+\sigma)^2}{a}\left[1+\dfrac{\mc{Y}}{(a+\sigma)^2}\right]} \label{33049}\\
\dot{\mc{Y}} &=& 2(a+\sigma)\mc{X}-a\mc{Y}+\mc{X}^2 \label{33050}
\en
Linear stability analysis in this case about the fixed point $(0,0)$ shows that it is a center for
\beq
a = \f{1-\sigma}{2}\pm\sqrt{\f{(1-\sigma)^2}{4}-\sigma} \label{33051}
\enq
This curve is shown in Fig. {\ref{f:4}}.
The interior region has the unstable focus and hence in this part of the parameter space there is a limit cycle. The range of $\sigma$ is limited to $0<\sigma\leq3-2\sqrt{2}$ and `$a$' lies between $0$ and $1$. Clearly the limit cycle at $a=1$ and $\sigma\neq 0$ is ruled out as expected.
\subsection{Summarizing...}
The above example serves the purpose of establishing our main point that the existence of a limit cycle would mean a flow equation of the form written down in Eq\eqref{eq2} and providing a method (although perturbative) of calculating the function f(A). In the process, we note the following facts:
\begin{itemize}
\item If a stable limit cycle exists, then there must exist an unstable focus.
\item If there are not enough parameters in the system to tune the focus to a center, then the linear terms in the system can be supplied with variable coefficients which may be tuned to yield a center. Perturbation theory can be carried out around this center.
\end{itemize}
It may very well be that there exists a family of limit cycles surrounding a focus. Such a case is naturally taken care of in our methodology, as the RG flow equation for amplitude --- $dA/d\tau=f(A)$--- would result in more than one fixed points: If there actually are $N$ limit cycles, then $f(A)=0$ will have $N$ positive real roots.
\section{Non-autonomous systems}

Having clearly illustrated how to distinguish between focus, center and limit cycle in a two-dimensional autonomous dynamical system, we now examine how our methodology fares in somewhat more complicated cases of non-autonomous systems and systems with time-delay.
\subsection{A damped driven ocsillator}
 We begin with a damped driven oscillator
\be
\ddot{x}+\omega^2x+k\dot{x}=F\cos\Omega t\label{34001a}
\en
 which we write as,
\be
\ddot{x}+\Omega^2x=-k\dot{x}+F\cos\Omega t+(\Omega^2-\omega^2)x.\label{34001}
\en
We treat $k$, $F$ and $\Omega^2-\omega^2$ as small to perturb about a center ($k = F =\Omega^2-\omega^2=0$ ).  Accordingly, proceeding as explained earlier, to the first order in all these small parameters, we obtain:
\be
\frac{dA}{d\tau}=-\frac{kA}{2}-\frac{F\sin\Theta}{2\Omega};\quad
\frac{d\Theta}{d\tau}=-\frac{F\cos\Theta}{2\Omega A}+\Delta\omega\label{34002}
\en
where $\Delta\omega\equiv\omega-\Omega$. Since, $\Omega$ is maintained externally, it cannot change, implying $d\Theta/d\tau=0$. Also, existence of fixed point requires $dA/d\tau=0$.
Therefore, the fixed point corresponds to the amplitude $A=F/[k^2+4(\Delta\omega)^2]^{1/2}$ and the phase $\Theta=\tan^{-1}[-{k}/{2(\Delta\omega)}]$.
This is exactly in accordance with the literature of forced oscillators.
The stable non-zero fixed point in the evolution of $A$ corresponds to a limit cycle in accordance with what we have claimed has to happen.
\subsection{Time delay equations}
Now let us consider oscillators with time delay.
A linear form of such an oscillator satisfies the differential equation:
\be
\ddot{x}(t)+\omega^2x(t)+\v x(t-t_d)=0\label{34003}
\en
Here we treat $\v$ as small and consider the perturbation of the center $(\v=0)$.
The RG flow equations upto $\mc{O}(\v)$ are found to be:
\be
\frac{dA}{d\tau}=\v\frac{\sin\omega t_d}{2\omega};\quad
\frac{d\Theta}{d\tau}=\v\frac{\cos\omega t_d}{2\omega} \label{34004}
\en
For a periodic orbit (center), our claim is: ${dA}/{d\tau}$ should be trivially zero; and this yields $\omega t_d =\pi$.
This may be compared with the exact result that equation \eqref{34003} exhibits oscillatory solution $A\exp(it\sqrt{\omega^2-\v})+c.c.$ when $t_d=\pi/\sqrt{(\omega^2-\v)}$.
The frequency of the periodic orbit is seen to be $\omega-{\v}/{2\omega}+\mathcal{O}(\v^2)$ in accordance with the expansion of the exact answer of equation \eqref{34003}.
\\
Similarly, one can study limit cycles too in weakly nonlinear time delayed systems with success \cite{ptpgoto}. The system we study next to illustrate that RG can be successfully implemented in such systems, is given by
\beq
\f{dx(t)}{dt}+\alpha x(t)+\beta x(t-t_d) = \lambda\left(x(t)-x^3(t)\right)\label{34005}
\enq
where $\alpha$, $\beta$ $\lambda$ and are constants, $\lambda$ being small. The LHS of Eq.\eqref{34005} constitutes the unperturbed system and the nonlinear terms in RHS will be treated as the perturbation. We proceed as usual with a naive expansion of the form $x(t)=x_0(t)+\lambda x_1(t)+\lambda^2 x_2(t)+\cdots$. At zeroth order, we have
\be
\f{dx_0(t)}{dt}+\alpha x_0(t)+\beta x_0(t-t_d)=0 \label{34006}
\en
It is easy to see analytically that the above equation has an oscillatory solution given the following condition is satisfied.
\beq
t_d = \f{\cos^{-1}\left(\alpha/\beta\right)}{\sqrt{\beta^2-\alpha^2}} \label{34007}
\enq
Restricting ourselves to cases where the above condition holds we find
\beq
x_0(t) = A_0\cos \omega t\label{34008}
\enq
where $\omega = \sqrt{\beta^2-\alpha^2}$ ; $\beta>\alpha$. Further we find: $\beta\sin\omega r =\omega$ and $\beta\cos\omega r =-\alpha$.
At order $\mc{O}(\lambda^1)$, using Eq. \eqref{34008} we have
\be
\f{dx_1(t)}{dt}+\alpha x_1(t)+\beta x_1(t-t_d) &=& x_0(t) - x_0^3(t) \nonumber\\
&=& \left(A_0-\f{3 A_0^3}{4}\right)\cos\omega t -\f{A_0^3}{4} \cos 3\omega t \label{34009}
\en
A little bit of algebra yields the solution for $x_1(t)$ which reads
\beq
x_1(t) = \f{1+\alpha t_d}{\left(1+\alpha t_d\right)^2+(\omega t_d)^2} t\cos \omega t + \f{\omega t_d}{\left(1+\alpha t_d\right)^2+(\omega t_d)^2} t\sin \omega t \label{34010}
\enq
To $\mc{O}(\lambda^1)$, thus, we have
\be
x(t) = A_0\cos \omega t+\lambda\f{1+\alpha t_d}{\left(1+\alpha t_d\right)^2+(\omega t_d)^2} t\cos \omega t + \lambda\f{\omega t_d}{\left(1+\alpha t_d\right)^2+(\omega t_d)^2} t\sin \omega t \label{34011}
\en
From this point onwards proceeding as described earlier we arrive at the RG flow equation right upto $\mc{O}(\lambda^1)$, given by
\be
\f{dA}{d\tau} = \f{\lambda A (1+\alpha t_d)}{\left(1+\alpha t_d\right)^2+(\omega t_d)^2}\left[1-\f{3}{4}A^2\right]; \quad \f{d\theta}{d\tau} = \f{\lambda A (\omega t_d)}{\left(1+\alpha t_d\right)^2+(\omega t_d)^2}\left[1-\f{3}{4}A^2\right]\label{34012}
\en
In accordance to our classification scheme we can immediately conclude that the system given by Eq.\eqref{34005} exhibits limit cycle oscillations. Amplitude of the limit cycle is given by the stable fixed point $A^2 = 4/3$.
\section{Advantage Over Linear Stability Analysis}

Before we conclude, let us witness how useful this RG technique is when one deals with the subtle cases of centers in nonlinear dynamical systems. It is very well known that linearized version of a nonlinear dynamical system may not reproduce qualitatively correct picture of the phase portrait near a fixed point. We now showcase the fact that while linearization of a certain nonlinear dynamical system wrongly establishes a fixed point as center (which originally is a spiral node), our methodology gives correct result. Consider the following dynamical system:
\begin{subequations}
\begin{equation}
\dot{x}=-y+\varepsilon ax(x^2+y^2 )
\end{equation}
\begin{equation}
\dot{y}= + x + \varepsilon ay(x^2+y^2 )
\end{equation}
\label{linstab}
\end{subequations}
Here, $\varepsilon$ is a small positive parameter that facilitates a trial perturbative solution of the form: $x(t)= x_0+\varepsilon x_1+\varepsilon^2 x_2+\cdots$ . Linear stability analysis would show that the fixed point $(0,0)$ is a center for all $a$. It can however be easily shown \citep{nldstr} by making use of polar coordinates, in system (\ref{linstab}), the origin is a stable spiral when $a < 0$ and an unstable spiral for positive $a$. Now, applying the RG methodology prescribed in this paper, one arrives at the following flow equations, upto $O(\varepsilon^2)$:
\begin{equation}
dA/d\tau=aA^3;\quad
d\theta/d\tau=0
\end{equation}
One immediately notes that in accordance with our scheme of classifying focus and center, from the above flow equations, one can easily extract the correct information regarding the nature of the fixed point in system (\ref{linstab}): if $a = 0$, $dA/d\tau$ = 0 $\forall A$ implying that the origin is a center; whereas if $a \ne 0$, $dA/d\tau=0\, iff\, A = 0$, making the origin a focus.
\\
One may recall that the fixed point given by expression (\ref{33021}) for the glycolytic oscillator defined by equations (\ref{33019}) and (\ref{33020}) was found to be a center on curve: $2a=\sqrt{1+8b^2}-(1+2b^2)$. However, this is a result of linear stability analysis where, by dint of very nature of the analysis technique, the fixed point is shielded from full bombardment of non-linear terms. For a specific value of $(a,b)=(1/8,\sqrt{3/8})$ lying on the curve, one obtains the flow equation [setting $\delta=0$ in relation (\ref{33030})]: $\pa A/\pa \tau=-3A^3/8$ (to the lowest order). One shouldn't be confused to observe that in accordance with the form of this flow equation, our prescription claims $(a,b)=(1/8,\sqrt{3/8})$ is actually a focus and {\it not} a center. Numerical simulations easily confirm this fact.
\section{Conclusions}

     To conclude, we again emphasize that this paper introduces a simple yet powerful methodology --- based on perturbative renormalization group theory --- of identifying and classifying a periodic solution (limit-cycle or orbit around center) in various types of two-dimensional nonlinear dynamical system. This very technique can also distinguish between a focus and a center.
The different types of two-dimesional systems that can be handled using this methodology include not only simpler autonomous systems but also forced non-autonomous systems and time-delayed systems..
Also, it has been shown that our technique yields the correct nature of the fixed point of a nonlinear dynamical system when the linearization about it gives a completely wrong idea regarding its true nature.\\
Given the inter-disciplinary nature of the subject of nonlinear dynamics and the wide research interest in investigating periodic solutions, our method should be of direct interest and practical use to researchers across scientific disciplines.
\acknowledgements
SC acknowledges academic and financial support from NBIA, Copenhagen and post-doctoral FNU research grant no. 505100-50 - 30,168 by Danish Research Council.
\bibliographystyle{plainnat}

\end{document}